\documentclass[aps, prx,superscriptaddress, twocolumn,secnumarabic,nobalancelastpage,nofootinbib]{revtex4-1} %twocolumn,
\usepackage{graphicx}
\usepackage{amsmath,amssymb}
\usepackage[utf8]{inputenc}
\usepackage[T1]{fontenc}
\usepackage{lmodern}
\usepackage{natbib}
\usepackage{hyperref}
\usepackage{xcolor}
\hypersetup{colorlinks=true,citecolor={blue},linkcolor={blue},urlcolor={blue}}
\usepackage{units}
\usepackage[english]{babel}
\usepackage[normalem]{ulem}
\usepackage{cancel}
\usepackage{upgreek}
\usepackage{wasysym}
\usepackage{braket} %For braket notation 
\usepackage{placeins} % Use for FloatBarriers to force figures beeing drawn before new section
\usepackage{float}
\usepackage{soul} %For strikethrough \st
\newcommand{\vect}[1]{\mathbf{#1}}
\makeindex

\begin{document}

\definecolor{orange}{RGB}{255, 69, 0}
\definecolor{green}{RGB}{26,148,49}
\setstcolor{red}
\newcommand{\hs}[1]{\textcolor{purple}{#1}}
\newcommand{\jdt}[1]{\textcolor{green}{#1}}
\newcommand{\sa}[1]{\textcolor{red}{#1}}

\title{Lotka-Volterra population dynamics in coherent and tunable oscillators of trapped polariton condensates}

\date{\today}

\author{J. D. T\"opfer}
\email{J.D.Toepfer@soton.ac.uk}
\affiliation{School of Physics and Astronomy, University of Southampton, Southampton, SO171BJ, United Kingdom}

\author{H. Sigurdsson}
\affiliation{School of Physics and Astronomy, University of Southampton, Southampton, SO171BJ, United Kingdom}
\affiliation{Skolkovo Institute of Science and Technology, Moscow, Bolshoy Boulevard 30, bld. 1, 121205, Russia}

\author{S. Alyatkin}
\affiliation{Skolkovo Institute of Science and Technology, Moscow, Bolshoy Boulevard 30, bld. 1, 121205, Russia}

\author{P. G. Lagoudakis}
\email{Pavlos.Lagoudakis@soton.ac.uk}
\affiliation{School of Physics and Astronomy, University of Southampton, Southampton, SO171BJ, United Kingdom}
\affiliation{Skolkovo Institute of Science and Technology, Moscow, Bolshoy Boulevard 30, bld. 1, 121205, Russia}

\renewcommand{\abstractname}{} %removes writing of 'Abstract'
%%%%%%%%%%%%%%%%%%%%%%%%%%%%%%%%%%%%%%%%%
%%%%%%%%%%%%%%%%%%%%%%%%%%%%%%%%%%%%%%%%%
\begin{abstract}
We demonstrate a regime in which matter-wave condensates of exciton-polaritons trapped in an elliptically shaped two-dimensional potential appear as a coherent mixture of ground and first-excited state of the quantum harmonic oscillator. This system resembles an optically controllable two-level system and produces near terahertz harmonic oscillations of the condensate's center of mass along the major axis of the elliptical trapping potential. The population ratio between the two trap levels is tunable through the excitation laser power and is shown to follow Lotka-Volterra dynamics. We demonstrate coherence formation between two spatially displaced trapped condensate oscillators - the polaritonic analogue of Huygen's clock synchronization for coupled condensate oscillators.
\end{abstract}
\maketitle
%%%%%%%%%%%%%%%%%%%%%%%%%%%%%%%%%%%%%%%%%
%%%%%%%%%%%%%%%%%%%%%%%%%%%%%%%%%%%%%%%%%
\section{Introduction}

The competitive Lotka–Volterra model describes coupled autonomous systems (or {\it species} in terms of predator-prey settings) which fight over a common resource. Such competition can be found in brain activity~\cite{Rabinovich_PLOS2008}, economics~\cite{goodwin1982growth}, and laser systems described by Maxwell-Bloch equations where different modes compete over the laser gain~\cite{Hacinliyan_NonlDyn2010}. However, dissipative Bose-Einstein condensates can also display similar competative population dynamics given the right setting. The purest form of condensation in the quantum sense refers to an equilibrium bosonic gas macroscopically occupying the system ground state while other levels deplete~\cite{Mueller_PRA2006}. When non-Hermiticity is added to the picture, which is the case for photon and polariton condensates, this macroscopic occupation is no longer exclusive to just the ground state but instead other available higher energy levels might become macroscopically occupied, depending on their internal losses and coupling to other levels leading to competition over the condensate gain.

While many-particle two-level systems such as lasers and condensates can give great insight to fundamental physics of coupled systems, they also form the solid-state implementation of the qubit, like superconducting circuits, for quantum information processing. Indeed, the possibility of using Bose-Einstein condensates for quantum computation was recently proposed~\cite{Byrnes_PRA2012}, followed by some exciting work focused on exciton-polariton condensates due to their easy optical manipulation~\cite{Demirchyan_PRL2014, Cuevas_SciAdv2018, Ghosh_npj2020, xue2019splitring}. Moreover, non-Hermitian two-level systems of polariton condensates could potentially model the open Bose-Hubbard dimer~\cite{Lledo_PRB2019}, $\mathcal{PT}$-symmetric physics~\cite{Bender_PRL2007, Chestnov_SciRep2016}, and nonlinear Josephson effects such as macroscopic self-trapping~\cite{Ostrovskaya_PRA2000}. Recently, they have been theoretically proposed as clock generators under resonant laser excitation~\cite{PhysRevB.101.115418}, and to realize propagating parity domain walls in extended condensates~\cite{PhysRevB.92.195409}. It is therefore of quite some interest to be able to generate and control macroscopically occupied dissipative two-level systems to investigate the aforementioned possibilities. Such control also opens new perspectives to study synchronization in extended two-level bosonic systems where adjacent traps possess different level occupation.

Our study is then motivated by the following question: Can a two-level dissipative condensate system be designed with tunable energy spacing and competitive population dynamics which determine the ratio of particles between the excited state and the ground state? In other words, can the condensate Bloch vector be tuned to have arbitrary polar angle on the Bloch sphere and precession frequency? Here, we answer this question in the affirmative. We show experimentally and theoretically that a non-resonantly excited exciton-polariton condensate, possessing two energy levels from its laser induced trap~\cite{Askitopoulos_2013PRB}, displays a gradual change in the relative level occupation number as a function of excitation power while the laser geometry determines the level splitting. The condensate population dynamics are found to follow exactly the competitive Lotka-Volterra equations due to cross-saturation effects coupling the levels together. Our modeling provides a transparent description of the physics at play and strengthen the connection between dissipative polariton condensates and laser systems. We additionally demonstrate the scalability of our system by measuring the coherent coupling between two two-level condensates by using two distinct laser beams. There, we observe that the dissipative nature of their tunneling current synchronizes the systems to maximizes their mutual gain.

%%%%%%%%%%%%%%%%%%%%%%%%%%%%%%%%%%%%%%%%%
\section{Results}
The dissipative nature of polariton condensates requires continuous pumping to compensate for (mainly optical) losses and maintain the condensate beyond its intrinsic $\sim \mathrm{ps}$ lifetime. Optical excitation of polariton condensates by means of a non-resonant laser beam is interlinked with the generation of an incoherent reservoir of excitons, which acts as both, optical gain and a repulsive potential for polaritons. The spatial distribution of this particle interaction-induced potential is inherently linked to the spatial geometry of the excitation pump profile. Design of potential landscapes by means of optical pump laser beam-shaping is at the heart of research for polaritonic devices and applications. While the use of widely-spaced tightly-focused pump laser beams (spot size $\approx 2\;\mathrm{\upmu m}$)  leads to the excitation of high-energy and radially expanding condensates~\cite{PhysRevX.6.031032,topfer2020time}, on the contrary, the use of closely-spaced pump beam geometries can show condensation in low-energy and spatially confined polariton modes~\cite{tosi2012sculpting, PhysRevLett.110.186403, PhysRevX.5.031002}. In particular, it has been shown that a non-resonant circular or elliptical annular pump profile creates a near-parabolic potential landscape for polaritons and facilitates trapped polariton condensation in states approximately given by the eigenstates $\Psi_{nm}$ of the two-dimensional quantum harmonic oscillator (HO)~\cite{PhysRevB.92.035305,PhysRevB.97.235303}. Change in experimental parameters, such as the diameter or the pump power of the annular excitation profile, modifies the system's resonance conditions and allows for condensation in different HO states characterized by their indices $n$ and $m$.

\begin{figure}[!t]
	\center
	\includegraphics[]{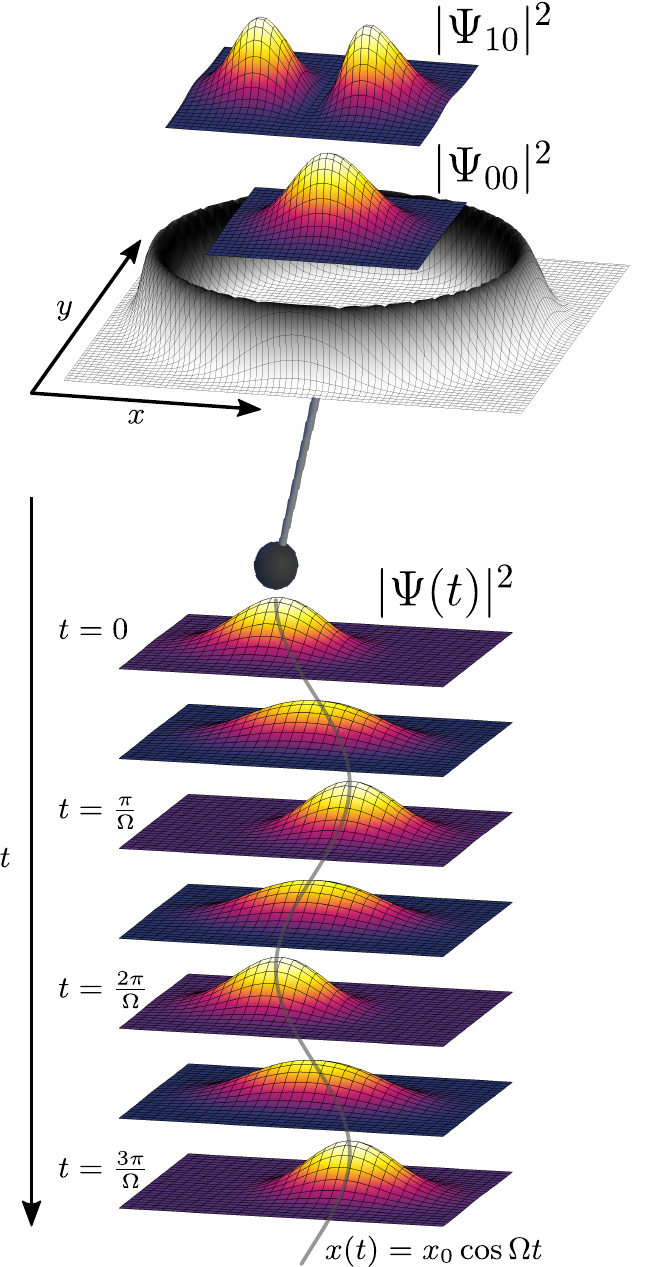}
	\caption{Schematic illustration of harmonic oscillations of a polariton condensate in an optically generated elliptical trapping potential. Dual-mode condensation into the two energetically lowest modes $\Psi_{00}$ and $\Psi_{10}$ with energy splitting $\hbar \Omega$ leads to an oscillatory motion of the condensate's center of mass $\braket{x}\propto \cos{(\Omega t)}$ in analogy to the harmonic motion of a mechanical pendulum for small displacement angles.}
	\label{Fig1}
\end{figure}

In the following, we demonstrate polariton condensation in an elliptically-shaped potential favoring condensation in the lowest two HO modes. Nonlinear coupling between the two energetically split modes (cross-saturation effects) is shown to yield a limit cycle state. This non-stationary state involves linear harmonic oscillations of the condensates center of mass $\braket{x}$ in analogy to the harmonic oscillations of a mechanical pendulum for small angles of displacement (see Fig.~\ref{Fig1}).
%
%%%%%%%%%%%%%%%%%%%%%%%%%%%%%%%%%%%%%%%%%
%%%%%%%%%%%%%%%%%%%%%%%%%%%%%%%%%%%%%%%%%
\subsection{Competition of two condensate modes}
\begin{figure}[!t]
	\center
	\includegraphics[]{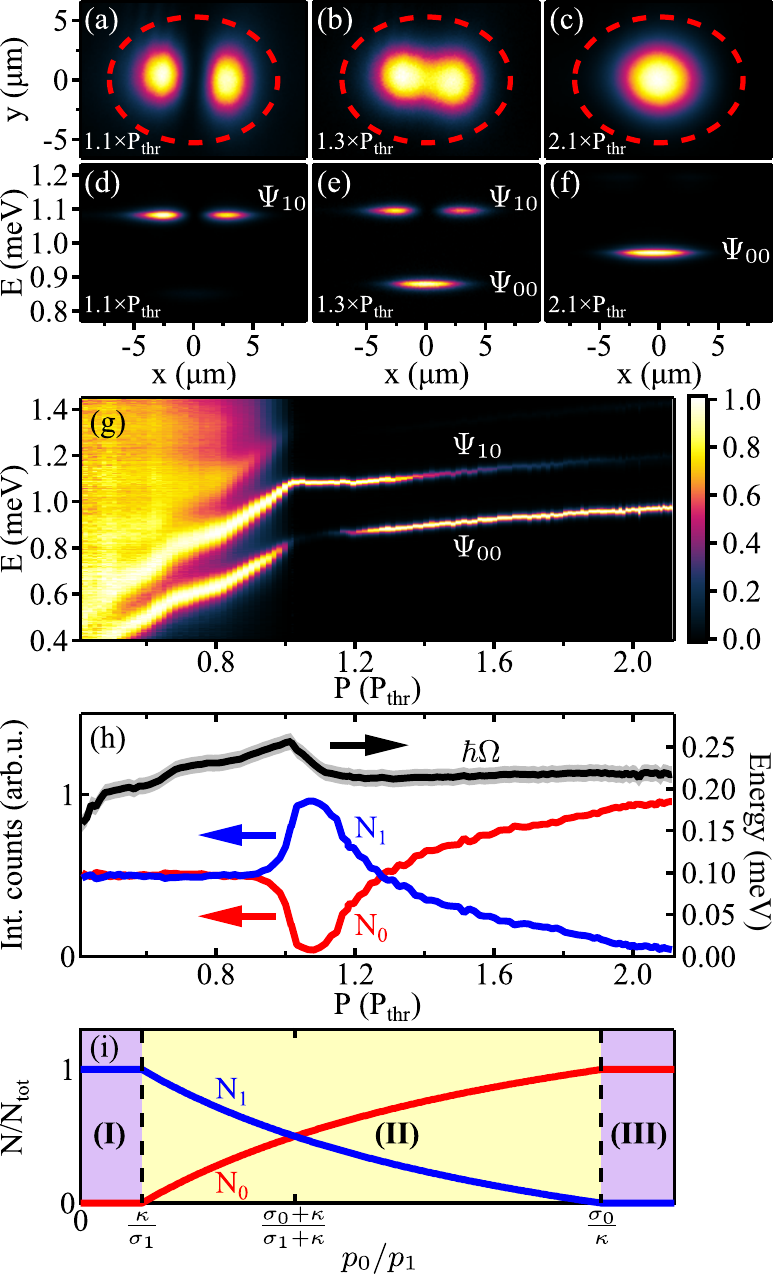}
	\caption{Two-mode competition in a trapped polariton condensate. (a-c) Real space and (d-f) energy-resolved real space photoluminescence along the major axis ($y=0$) of the elliptical annular pump beam (red-dashed line) for varying excitation pump power $P$ above condensation threshold. (g) Integrated spectra for increasing excitation pump power $P$. (h) Spectral weights of the ground-state mode (red) and the first excited state (blue) and their energy splitting $\hbar \Omega$ (black). The experimental spectral resolution ($0.02\;\mathrm{meV}$) is illustrated in gray-shaded. Energy scale in (d-g) is scaled as the blueshift with respect to the lower polariton ground state mode below threshold. (i) Relative mode occupations $N_{0,1}/N_{\mathrm{tot}}$ with $N_{\mathrm{tot}} = (N_0+N_1)$ for two weakly coupled modes described in Eqs.~\eqref{Eq.SingleTrap_a} and \eqref{Eq.SingleTrap_b} as a function of their relative net gain $p_0/p_1$.}
	\label{Fig2}
\end{figure}

By shaping the non-resonant continuous wave excitation laser profile into an elliptical annulus (see Methods) with diameters of $14.2\;\mathrm{\upmu m}$ and $10.6\;\mathrm{\upmu m}$ in $x$- and $y$-direction, respectively, we lift the degeneracy of the two excited HO states $\Psi_{10}$ and $\Psi_{01}$~\cite{PhysRevB.97.235303}. Here, the trap's elongation in x-direction leads to a larger net gain for the $\Psi_{10}$ state as for the $\Psi_{01}$ state, thus, favoring condensation into the former state. In Fig.~\ref{Fig2}(a) we show the measured real-space polariton photoluminescence (PL) for the system excited with pump power $P=1.1P_{\mathrm{thr}}$, where $P_{\mathrm{thr}}$ is the system's condensation threshold pump power. We confirm single-mode condensation in the $\Psi_{10}$ state by spectrally resolving the emission along the major axis of the elliptical trap ($y=0$) as illustrated in Fig.~\ref{Fig2}(d). The observed blueshift of $\approx 1.1\;\mathrm{meV}$ above the lower polariton ground state energy is a result of the optically induced background (confinement) potential. By increasing the excitation pump power to  $P=2.1P_{\mathrm{thr}}$ we observe a transition of the system from single-mode excited state $\Psi_{10}$ occupation to single-mode ground state $\Psi_{00}$ occupation as shown in Figs.~\ref{Fig2}(c) and~\ref{Fig2}(f). Interestingly, in the transition region, i.e. $1.1P_{\mathrm{thr}}<P<2.1P_{\mathrm{thr}}$, we observe the coexistence of both modes. Such dual-mode operation with approximately equal occupation of both modes is realized at $P=1.3P_{\mathrm{thr}}$ and shown in Figs.~\ref{Fig2}(b) and~\ref{Fig2}(e). The continuous transition of excited state to ground state condensation with increasing pump power $P$ is illustrated in Figs.~\ref{Fig2}(g) and~\ref{Fig2}(h) showing the (spatially integrated) spectra and normalised mode emission intensities (blue and red curves), respectively. We point out that the mode linewidths for the spectra illustrated in Figs.~\ref{Fig2}(d-g) are limited by the spectral resolution $\Delta E \approx 0.02\;\mathrm{meV}$ of our experimental setup. We separately estimate a linewidth of $\approx 0.004\;\mathrm{meV}$ for the ground state condensate at $P=2.1P_{\mathrm{thr}}$ from analysis of the condensate's auto-correlation function (see Appendix~\ref{Appendix_section_CoherenceTime}).

While the energy levels of both modes $E_{00}$ and $E_{10}$ experience a continuous blueshift with increasing excitation pump power $P>1.1P_{\mathrm{thr}}$ we find that - within our experimental resolution - there is no change in energy splitting $\hbar \Omega = E_{10}-E_{00}  \approx 0.22\;\mathrm{meV}$ (see black line in Fig.~\ref{Fig2}(h)). Because the dominant contribution to the energy splitting $\hbar \Omega$ between ground and first excited state is given by the optically induced confinement potential, we conclude that there is no significant change in the potential landscape for increasing pump power $P$ above condensation threshold. 
Motivated by our experimental results we describe the condensate wave function $\Psi(\vect{r},t)$ in a two-mode model
\begin{equation}  \label{Eq.TwoModeModel}
\Psi(\vect{r},t)=\psi_0(t)\Psi_{00}(\vect{r}) + \psi_1(t)\Psi_{10}(\vect{r}),
\end{equation} 
where $\Psi_{00,10}$ represent the orthogonal and normalized spatial field distributions of the system's ground- and first-excited state. The complex-valued envelopes $\psi_{0,1}$ resemble occupation numbers $N_{0,1}=|\psi_{0,1}|^2 $ and phases $\theta_{0,1}=\arg(\psi_{0,1})$ of each mode, respectively.
One can derive the nonlinear-coupled two-level system (see Appendix~\ref{Appendix_Theory})
\begin{equation} 
\begin{aligned}[b]
i \dot{\psi}_0 &=  \left[ \omega_0 + i p_0 + (\alpha_0 - i \sigma_0) |\psi_0|^2 +  (\beta - i \kappa) |\psi_1|^2  \right] \psi_0, \\
i \dot{\psi}_1 &=  \left[ \omega_1 + i p_1 + (\alpha_1 - i \sigma_1) |\psi_1|^2 +  (\beta - i \kappa) |\psi_0|^2  \right] \psi_1. \label{Eq.complexEquations}
\end{aligned} 
\end{equation}
Here, for each mode $j=0,1$ the real-valued parameters denote eigenfrequency $\omega_j$, net gain $p_j$, self- and cross- nonlinearities $\alpha_j$ and $\beta$, as well as the self- and cross-saturation terms $\sigma_j $ and $\kappa$. Equations~\eqref{Eq.complexEquations} can analogously be rewritten into four coupled differential equations for the occupation numbers $N_{0,1}$ and phases $\theta_{0,1}$ of each mode, i.e.
\begin{subequations}
\begin{align}
\dot{N}_{0} &=  2\cdot \left( p_{0} - \sigma_{0} N_{0} - \kappa N_{1} \right) N_{0}, \label{Eq.SingleTrap_a} \\
\dot{N}_{1} &=  2\cdot \left( p_{1} - \sigma_{1} N_{1} - \kappa N_{0} \right) N_{1}, \label{Eq.SingleTrap_b} \\
\dot{\theta}_{0} &=   \omega_{0} + \alpha_{0} N_{0} + \beta N_{1},  \label{Eq.SingleTrap_c}  \\
\dot{\theta}_{1} &=   \omega_{1} + \alpha_{1} N_{1} + \beta N_{0}.  \label{Eq.SingleTrap_d} 
\end{align} 
\end{subequations}
Equations~\eqref{Eq.SingleTrap_a} and \eqref{Eq.SingleTrap_b}, which are decoupled from the phase dynamics, have previously been theoretically studied in the context of mode competition in polariton condensates~\cite{PhysRevB.78.035319,PhysRevE.101.012207}. They represent the competitive Lotka-Volterra equations, and as such describe competition between two interacting species populations~\cite{murray2007mathematical}, as well as the competition between two laser modes~\cite{sargent1974}. The relative strength of competition between the two modes sharing the same gain medium is given by the parameter $C=\kappa^2 / \sigma_1 \sigma_2$. While in the so-called strong coupling regime ($C>1$) cross-saturation effects dominate the system and one species will quench the population of the second entity, in the following, we restrict our analysis to the experimentally relevant weak coupling regime ($C<1$) which facilitates the coexistence of both species. Considering configurations where at least one mode is pumped above threshold, i.e. $p_{0,1}>0$ , there exist three non-trivial equilibrium points ($\dot{N}_{0,1}=0$) given by 
\begin{equation} 
\begin{aligned}[b]
&(\mathrm{I}) & &N_{0} =0, & &N_{1} =p_1/\sigma_1, \\
&(\mathrm{II}) & &N_{0} =\frac{\sigma_1 p_0 - \kappa p_1}{\sigma_0\sigma_1 - \kappa^2}, &  &N_{1} = \frac{\sigma_0 p_1 - \kappa p_0}{\sigma_0\sigma_1 - \kappa^2}, \\
&(\mathrm{III}) & &N_{0} =p_0/\sigma_0, &  &N_{1} =0. \label{Eq.EquilibriumPoints}
\end{aligned}
\end{equation} 
However, only one of these points can be stable for the same parameters and stability of each solution $(\mathrm{I}-\mathrm{III})$ is determined by the system's Jacobian which yields the conditions 
\begin{equation} 
\begin{aligned}[b]
&(\mathrm{I})   & &p_0 < p_1 \kappa / \sigma_1 , & &p_1 >                                0, \\
&(\mathrm{II})  & &p_0 > p_1 \kappa / \sigma_1 , & &p_1 > p_0 \kappa / \sigma_0, \\
&(\mathrm{III}) & &p_0 >                                0 , & &p_1 < p_0 \kappa / \sigma_0. \label{Eq.StabilityEquilibriumPoint}
\end{aligned}
\end{equation} 
While two of these points $(\mathrm{I},\mathrm{III})$ represent single-mode polariton emission, one point $(\mathrm{II})$ corresponds to dual-mode emission with energies $E_{0,1}=\hbar \dot{\theta}_{0,1}$ given by Eqs.~\eqref{Eq.SingleTrap_c} and \eqref{Eq.SingleTrap_d}.  In Fig.~\ref{Fig2}(i) we show the equilibrium occupation numbers $N_0$ and $N_1$ for increasing relative net gain $p_0/p_1$ between the two modes. The continuous transition from single-mode emission in the second mode (I) to single-mode emission in the first mode (III) is interleaved with a region of dual-mode emission (II). The physical reasons for the increase of $p_0/p_1$ in experiment as a function of laser power can be understood from two different mechanisms: First, as pump power is increased further above threshold the number of particles in the condensate increases which blueshifts the excited state out of the trap making it more and more lossy. Second, the growing number of particles in the reservoir enables polaritons to relax more efficiently in energy~\cite{Wouters_PRL2010, PhysRevB.92.035305} which makes $p_0$ grow faster than $p_1$.

\begin{figure*}[!t]
	\center
	\includegraphics[]{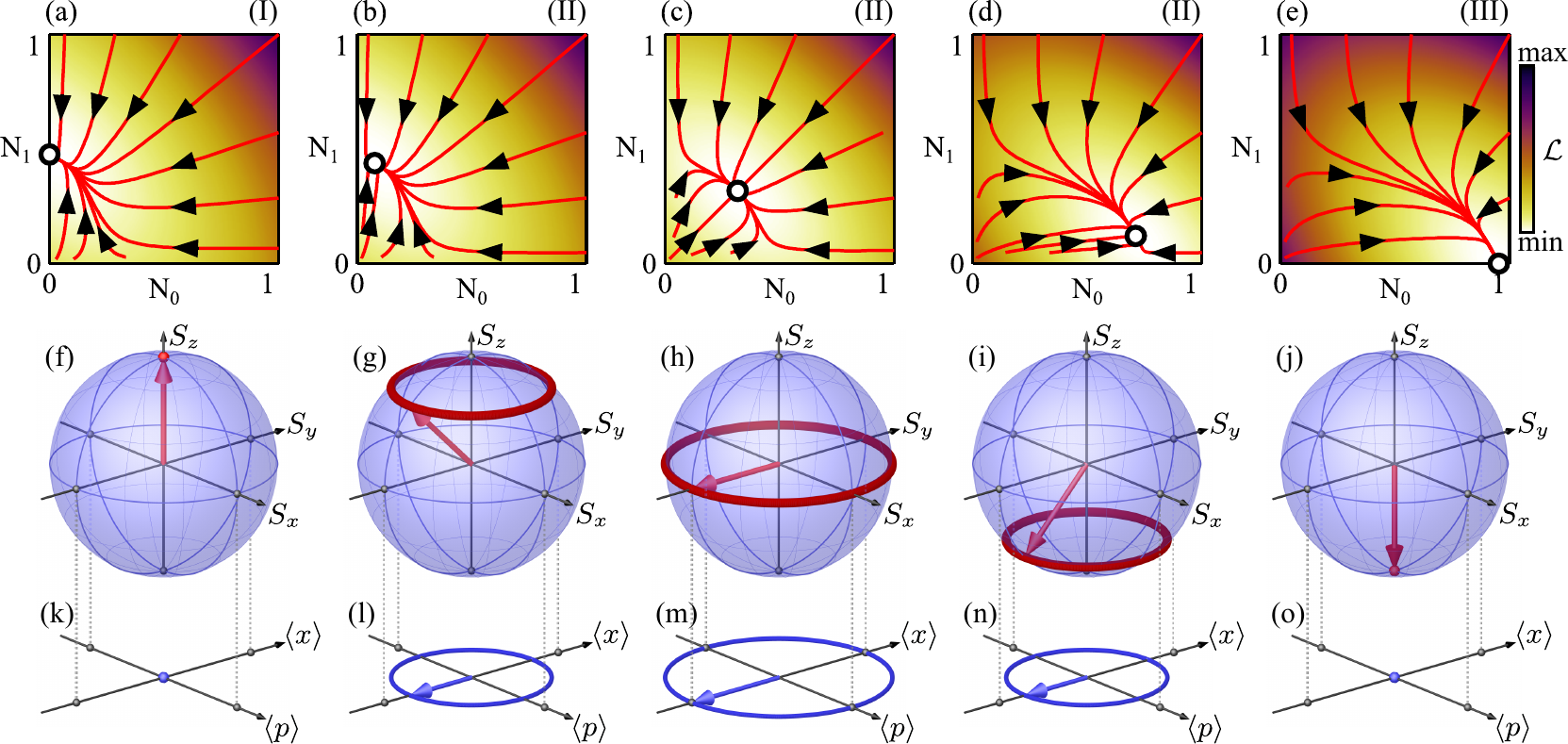}
	\caption{Numerical analysis of two weakly coupled condensate modes with occupation numbers $N_0$ and $N_1$ described in Eqs.~\eqref{Eq.SingleTrap_a} and \ref{Eq.SingleTrap_b}. (a-e) Phase portraits and Lyapunov potentials $\mathcal{L}$ (false-colorscale, Eq.~\eqref{Eq.LyapunovFunction}) for increasing relative net gain $p_0/p_1$ from left to right. Black circles depict fixed point attractors for vanishing occupation $N_0=0$ of the ground state (a), dual-mode operation $N_{0,1} > 0$ (b-d) and vanishing occupation $N_1=0$ of the excited state (e). (f-j) Representation of the intensity equilibrium points on the Bloch sphere spanned by the three-component vector $\vect{S}=(S_x,S_y,S_z)$ described in Eq.~\eqref{Eq.S_Vector}. (k-o) Projection of the orbits on the Bloch sphere (red) onto the  two-dimensional phase-space (blue) spanned by the condensate's center of mass $\langle x \rangle$ and momentum expectation value $\langle p \rangle$ illustrates the regimes in which the trapped condensate is stationary (k,o) and harmonically oscillating (l-n). Numerical parameters: $\sigma_0=\sigma_1=1\;\mathrm{ps}^{-1}$, $p_1=\kappa=0.5\;\mathrm{ps}^{-1}$ and $p_0= (0.25\;\mathrm{ps}^{-1},0.31\;\mathrm{ps}^{-1}, 0.5\;\mathrm{ps}^{-1},0.81\;\mathrm{ps}^{-1}, 1\;\mathrm{ps}^{-1})$ for (a-e) and (f-j), respectively.}
	\label{Fig_PhasePortraits}
\end{figure*}

We calculate the transient behavior of the competitive Lotka-Volterra system towards its fixed point attractors for all three scenarios (I-III) as shown in Figs.~\ref{Fig_PhasePortraits}(a-e). Here, for each configuration, we also plot the system's Lyapunov function $\mathcal{L}(N_0, N_1)$ in false-colorscale which can be written in quadratic form~\cite{macarthur1970species},
\begin{equation} 
\mathcal{L} = -2p_0 N_0 - 2p_1 N_1  
+ \sigma_0 N_0^2 +\sigma_1 N_1^2 +2 \kappa N_0 N_1. \label{Eq.LyapunovFunction}
\end{equation} 
We point out that the Lyapunov function $\mathcal{L}$ satisfies the condition $\dot{\mathcal{L}}\leq 0$ and that the system's fixed point attractors correspond to the minima of $\mathcal{L}(N_0,N_1)$ within its domain $\left\{N_0 \geq 0,  N_1 \geq 0\right\}$. We stress that the observed change in mode populations here is very different from the mode switching previously reported in~\cite{PhysRevB.97.045303} which was attributed to linear (deterministic) polariton physics where the laser parameters changed which quantum mode had the lowest condensation threshold.

The presented configurations with stable mode occupation numbers $N_{0,1}$ are solutions of the nonlinear coupled two-level system $\boldsymbol{\psi}(t) = (\psi_1(t),\psi_0(t))^T$ described in Eqs.~\eqref{Eq.complexEquations}. Hereby, the constant emission energy $E_{0,1}=\hbar \dot{\theta}_{0,1}$ of each mode is blue-shifted and determined in Eqs.~\eqref{Eq.SingleTrap_c} and \eqref{Eq.SingleTrap_d}. These two-level states $\boldsymbol{\psi}(t)$ describe orbits on the Bloch sphere spanned by the three-component vector $\vect{S}=(\boldsymbol{\psi}^{\dagger}\boldsymbol{\sigma}\boldsymbol{\psi})/(\boldsymbol{\psi}^{\dagger}\boldsymbol{\psi})$ with Pauli matrix vector written $\boldsymbol{\sigma}$. One can show that
\begin{equation} 
\begin{aligned}[b]
S_x(t) &= \frac{2 \sqrt{N_1 N_0}}{N_1 + N_0} \cos(\Omega t + \delta), \\
S_y(t) &= \frac{2 \sqrt{N_1 N_0}}{N_1 + N_0}  \sin(\Omega t + \delta), \\
S_z(t) &= \frac{N_1-N_0}{N_1 + N_0} . \label{Eq.S_Vector}
\end{aligned}
\end{equation} 
where $\delta$ is a stochastic phase-offset chosen during the spontaneous condensation event. In Figs.~\ref{Fig_PhasePortraits}(f-j) we illustrate the orbits corresponding to the intensity equilibrium points depicted in Figs.~\ref{Fig_PhasePortraits}(a-e). Condensate configurations with single-mode occupation in ground or excited state yield fixed points located at the poles of the Bloch-sphere as shown in (f) and (j). The periodic orbits in (g-i) for dual-mode operation are centered around the $S_z$-axis and their period is determined by the modes' energy splitting, i.e. $T=2\pi\Omega^{-1}$. Modification of the relative occupation numbers (as demonstrated in Fig.~\ref{Fig2}(h)) allows control over the position of the orbit's $S_z$-position. Furthermore, as we demonstrate in Appendix~\ref{Appendix_section_FrequencyTuning}, the period $T$ of these orbits is optically tuneable in the range of $10-20\;\mathrm{ps}$ through spatial modulation of the pump laser profile.

 %It should be underlined that condensation of cavity polaritons~\cite{Galbiati_PRL2012}, and quite recently for photons~\cite{Kurtscheid_Science2019}, is already possible in a two-level configuration where two spatially separated potential traps are brought together. The tunnelling current bridging the traps then splits the resonant energy levels of each potential into a bound and antibound state. Furthermore, spinor polariton condensates can also form a very fine two-level system in presence of cavity birefringence which splits the polarization (spin) levels of spinor polaritons~\cite{PhysRevX.5.031002}, or even by their own interactions~\cite{askitopoulos2020coherence}.
%
%
%%%%%%%%%%%%%%%%%%%%%%%%%%%%%%%%%%%%%%%%%
%%%%%%%%%%%%%%%%%%%%%%%%%%%%%%%%%%%%%%%%%
\subsection{Harmonic oscillations of a trapped condensate}

There is a direct relation between the periodic orbits appearing on the two-level system's Bloch-sphere described in Eq.~\eqref{Eq.S_Vector} and the harmonic oscillations of the condensate density $|\Psi(\vect{r},t)|^2$ illustrated in Fig.~\ref{Fig1}. In particular, the expectation values of the condensate's center of mass along the horizontal (major) axis in real space $\braket{x}$ and momentum space $\braket{p}$ are given by
\begin{subequations}
\begin{align} 
\braket{x}(t) &=  x_0 \cdot S_x(t), \label{Eq.density_x} \\
\braket{p}(t) &= p_0 \cdot S_y(t), \label{Eq.density_p}
\end{align} 
\end{subequations}
where the scaling factors $x_0 =\bra{\Psi_{00}}x\ket{\Psi_{10}}$ and $p_0 =\bra{\Psi_{00}}\hbar \partial_x \ket{\Psi_{10}}$ are dependent on the spatial distribution of the two competing modes. The trajectories of the two-level system on the Bloch sphere in Figs.~\ref{Fig_PhasePortraits}(f-j) can be projected onto the phase-plane spanned by space $\braket{x}$ and momentum $\braket{p}$ expectation values as illustrated in Figs.~\ref{Fig_PhasePortraits}(k-o). Here, single-mode condensation in either excited state (k) or ground state (o) both correspond to a stationary condensate center of mass located at $\braket{x}=\braket{p}=0$. The periodic orbits appearing in Figs.~\ref{Fig_PhasePortraits}(l-n) depict harmonic oscillations of the condensate's space $\braket{x}$ and momentum $\braket{p}$ expectation values in analogy to the phase-plane orbits of a mechanical undamped oscillator.

Although we cannot directly assess the condensate's ultrafast (ps-timescale) oscillations, we indirectly resolve them through time-correlation measurements. Hereby, we interfere the real-space polariton photoluminescence $\Psi(\vect{r},t)$ with a retro-reflected and time-shifted version $\Psi(-\vect{r},t+\tau)$ of itself and extract the system's first-order correlation function (see Methods)
\begin{equation} \label{Eq.CorrelationFunction}
g^{(1)}(-\vect{r},\vect{r};\tau)= \frac{\langle \Psi^*(\vect{r},t)  \Psi(-\vect{r},t+\tau) \rangle }{ \sqrt{ \langle |\Psi(\vect{r},t)|^2 \rangle \langle |\Psi(-\vect{r},t)|^2 \rangle}  },
\end{equation} 
where $\langle ... \rangle$ denotes time averaging. The coherence function (Eq.~\eqref{Eq.CorrelationFunction}) is a complex-valued and normalized measure for the first-order correlations between the two signals $\Psi(\vect{r},t)$ and $\Psi(-\vect{r},t+\tau)$. When the signals are composed of two frequency components with frequency-detuning $\Omega$ we expect periodic modulation of the coherence function $g^{(1)}(-\vect{r},\vect{r};\tau)$ at the same frequency $\Omega$ according to the Wiener-Khinchin theorem.

\begin{figure}[!t]
	\center
	\includegraphics[]{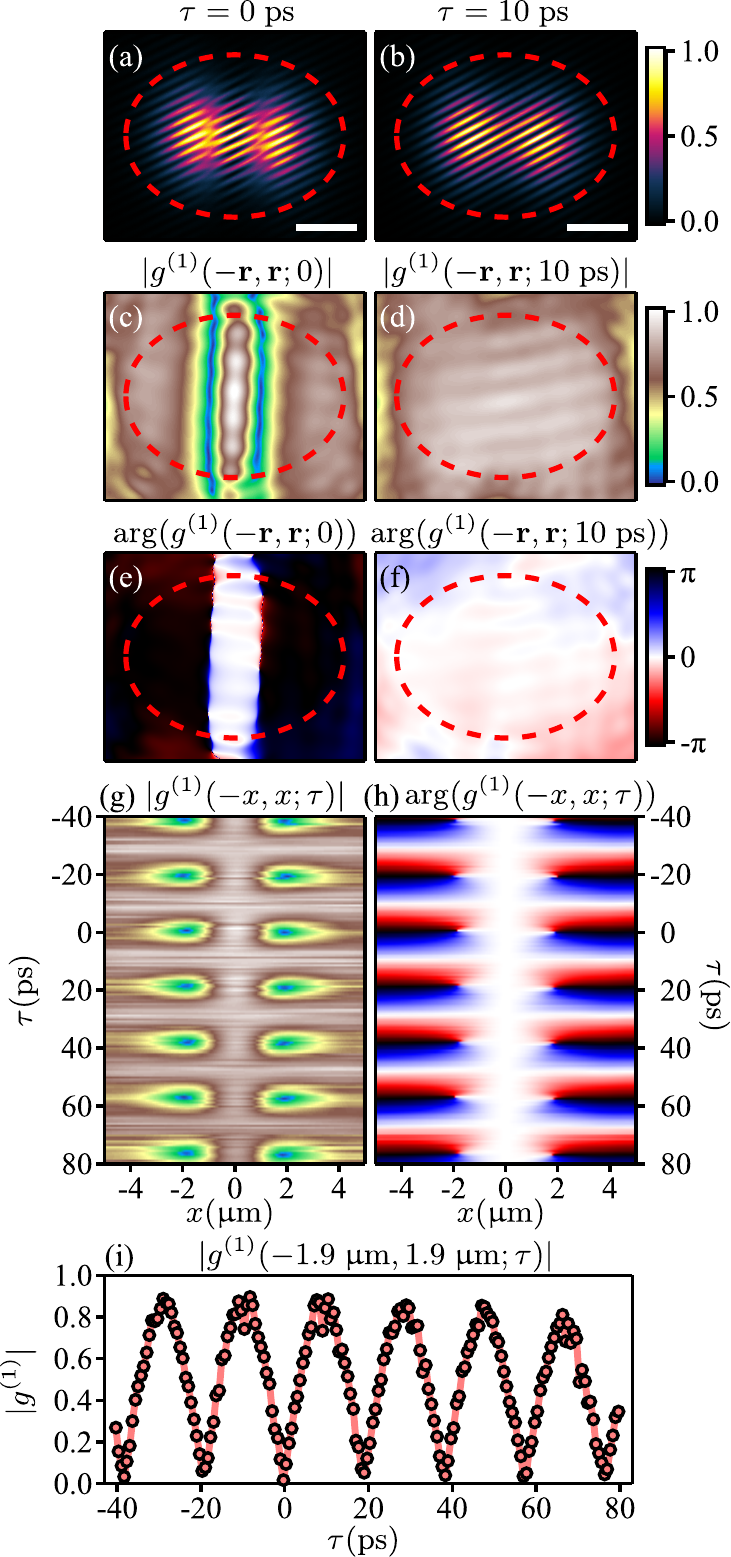}
	\caption{First-order correlations of a dual-mode polariton condensate in an elliptical trap. (a,b) Recorded interference of the real-space polariton photoluminescence $\Psi(\vect{r},t)$ and its retro-reflected and time-shifted version $\Psi(-\vect{r},t+\tau)$ for (a) $\tau=0$ and (b) $\tau=10\;\mathrm{ps}$. The annular pump profile is shown as a red-dashed ellipses. (c,d) Magnitude and (e,f) phase of the digitally reconstructed first-order coherence function $g^{(1)}(-\vect{r},\vect{r};\tau)$ for (c,e) $\tau=0$ and (d,f) $\tau=10\;\mathrm{ps}$, respectively. Temporal evolution of (g) magnitude and (h) phase of the coherence function $g^{(1)}(-x,x;\tau)$ along the horizontal slice $y=0$.  For better visibility an offset to the phase pattern has been applied for each time step $\tau$ to yield 0 phase at the center $r=0$. (i) Harmonic oscillation of the coherence function $|g^{(1)}(-x,x;\tau)|$ with maximum oscillation-amplitude at $x=1.9\;\mathrm{\upmu m}$. The condensate is pumped at $1.3$ times the condensation threshold. Scale bar in (a) corresponds to $4\;\mathrm{\upmu m}$ and applies to (a-f).}
	\label{Fig3}
\end{figure}

In the following we unravel the time-correlations of the trapped condensate system presented in Fig.~\ref{Fig2}(b,e) with approximately equal occupation of ground and first excited state. Recorded interference patterns and extracted modulus $|g^{(1)}|$ and argument $\arg (g^{(1)})$ of the correlation function are illustrated in Figs.~\ref{Fig3}(a,c,e) for no relative time-delay $\tau=0$ between the two signals. The equal-time correlations reveal vanishing coherence ($|g^{(1)}(-\vect{r},\vect{r};\tau =0)$) along the vertical lines $x \approx \pm  1.9\;\mathrm{\upmu m}$. This is a direct result of the mixing of two states with opposite parity and equal contribution to the condensate density, $N_0 |\Psi_{00}(\vect{r})|^2 = N_1 |\Psi_{10}(\vect{r})|^2$. On the contrary, at small distances $|x| \ll  1.9\;\mathrm{\upmu m}$, where the condensate density is mainly formed by particles in the ground state, we find a highly correlated signal $g^{(1)} \approx 1$ resembling the symmetric nature (even parity) of the ground state $\Psi_{00}$. At larger distances $|x| \gg 1.9\;\mathrm{\upmu m}$ we find an anti-correlated signal $g^{(1)} \approx -1$ as a result of the anti-symmetric field distribution (odd parity) of the excited state $\Psi_{10}$. The composition of even and odd parity states is a necessary condition for a non-vanishing oscillation amplitude $x_0 =\bra{\Psi_{00}}x\ket{\Psi_{10}}$ of the condensate's center of mass. \\

Introducing a temporal delay of half an oscillation period $\tau=\pi/\Omega \approx 10\;\mathrm{ps}$ we measure an interference pattern as illustrated in Fig.~\ref{Fig3}(b) with corresponding modulus and phase of the correlation function $g^{(1)}(-\vect{r},\vect{r};\tau =10\;\mathrm{ps})$ shown in Figs.~\ref{Fig3}(d) and (f). Here, we have applied a constant offset to the argument of the correlation function to yield 0 phase at the origin $\vect{r}=0$, i.e. we measure from a co-rotating reference frame. We find large correlation between spatially mirrored positions $g^{(1)}(-\vect{r},\vect{r};\tau =10\;\mathrm{ps}) \approx 1$ throughout the whole system in agreement with the condensate's spatial density oscillation. Further, in Figs.~\ref{Fig3}(g) and (h) we plot the extracted modulus and argument of the correlation function $g^{(1)}(-x,x;\tau)$ along the trap major axis $y=0$ for varying temporal delay $\tau$. Periodic disappearance and revival of coherence at $x=\pm 1.9\;\mathrm{\upmu m}$ (see Fig.~\ref{Fig3}(i)) with frequency $\Omega/2\pi \approx 52\;\mathrm{GHz}$ reveal the coherent nature of the present harmonic oscillations persisting over more than 20 oscillations ($\sim 400\;\mathrm{ps}$). 
%
%%%%%%%%%%%%%%%%%%%%%%%%%%%%%%%%%%%%%%%%%
%%%%%%%%%%%%%%%%%%%%%%%%%%%%%%%%%%%%%%%%%
\subsection{Synchronization of two polariton oscillators}
\begin{figure*}[!t]
	\center
	\includegraphics[]{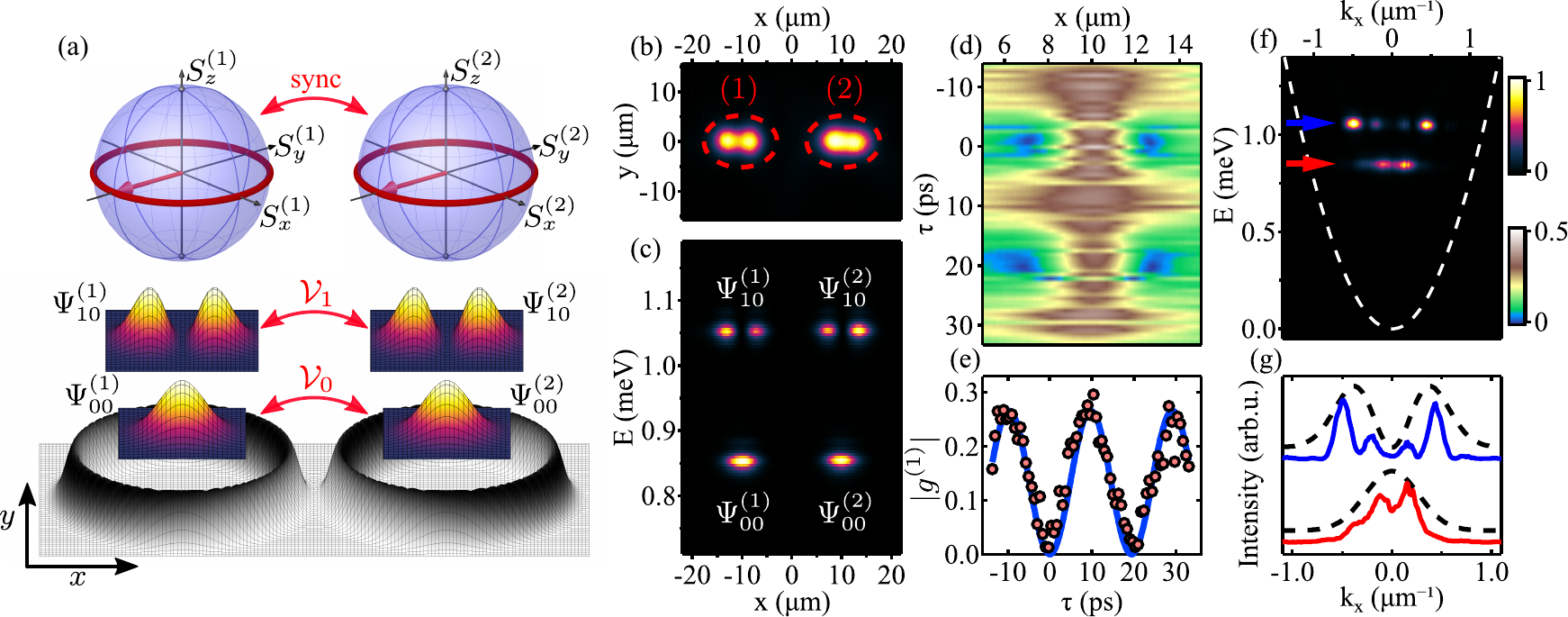}
	\caption{Synchronization of two polariton oscillators. (a) Schematic showing the coupling between two trapped and spatially separated dual-mode condensates and the resulting synchronization of their periodic orbits on the Bloch sphere. (b) Measured real-space condensate photoluminescence excited by two elliptical annular pump profiles (major axis $\approx 14.7\;\mathrm{\upmu m}$, minor axis $\approx 10.4\;\mathrm{\upmu m}$) shown as red-dashed lines. (c) Corresponding energy-resolved emission along symmetry axis $y=0$. (d) Modulus of the first-order coherence function $|g^{(1)}(-x,x;\tau)|$ showing correlations between the centers of the two trapped condensates. (e) Periodic disappearance and revival of coherence $|g^{(1)}(\tau)|$ at $x = 7\;\mathrm{\upmu m}$. (f) Energy-resolved momentum space emission along $k_y=0$. White dashed line depicts the lower polariton dispersion below condensation threshold. (g) Extracted momentum-space mode profiles for interfering ground states $\Psi^{(1,2)}_{00}$ (red) and excited states $\Psi^{(1,2)}_{10}$ (blue) at energies indicated by arrows in (g). Black dashed lines display the calculated mode profiles in the absence of interference, i.e. for incoherent modes.}
	\label{Fig4}
\end{figure*}

In a next step we investigate a system of two spatially separated polariton condensates, each of them representing a trapped (dual-mode) oscillator. As we schematically illustrate in Fig.~\ref{Fig4}(a) for small separation distances between the two traps a finite coupling rate $\mathcal{V}_{0,1}$ between the system's ground states $\Psi_{00}^{(1,2)}$ and excited states $\Psi_{10}^{(1,2)}$, where the superscript denotes different traps, leads to phase-locking of the condensates' periodic orbits and, thus, to synchronized oscillatory motion of the centers of mass $\braket{x^{(1,2)}}$ of both condensates. \\
For our experiment we generate two identical elliptical annular pump profiles with similar dimensions as those presented in Figs.~\ref{Fig2} and \ref{Fig3}. The two trap centers are displaced by $d\approx 20.7\;\mathrm{\upmu m}$ and excitation at pump power $P\approx 1.3P_\mathrm{thr}$ leads to approximately equal occupation of ground state $\Psi^{(1,2)}_{00}$ and excited state $\Psi^{(1,2)}_{10}$ in each trap $(1)$ and $(2)$, respectively. In Fig.~\ref{Fig4}(b) we show the measured real-space condensate PL and in Fig.~\ref{Fig4}(c) we plot the spectrally-resolved emission along the symmetry axis $y=0$. We note that - within the spectral resolution of our experiment - the energies of the ground and first excited modes are identical in both traps, where each trap has energy splitting $\hbar \Omega \approx 0.20\;\mathrm{meV}$. Coherence between the two spatially separated dual-mode condensates is confirmed by interferometric cross-correlation measurements. In particular, the measured modulus of the first-order correlation function $|g^{(1)}(-x,x;\tau)|$ along the symmetry axis $y=0$ is shown in Fig.~\ref{Fig4}(d). The illustrated spatial range $5\;\mathrm{\upmu m} < x < 15\;\mathrm{\upmu m}$ contains the mutual coherence between both traps centered at $x \approx \pm 10\;\mathrm{\upmu m}$. In analogy to our results for a single oscillating condensate displayed in Fig.~\ref{Fig3}(g) we observe spatially-distributed time-periodic modulation of the coherence function demonstrating synchronized spatial oscillations of both trapped condensates. A profile of the temporal coherence function with maximum oscillation amplitude at $x=7\;\mathrm{\upmu m}$ is displayed in Fig.~\ref{Fig4}(e) showing periodic disappearance and revival of coherence with frequency $\sim 50\;\mathrm{GHz}$. We point out that the measured coherence function displayed in Fig.~\ref{Fig4}(e) with vanishing correlation at $\tau = 0$ reveals phase-locking of the two oscillating condensate center of masses $\braket{x^{(1,2)}}$ with vanishing phase difference (in-phase synchronization).

Coherence between the two condensate traps and phase-locking of their respective oscillating centers of mass $\braket{x^{(1,2)}}$ is further confirmed by our measurement of the system's energy-resolved momentum space emission shown in Fig.~\ref{Fig4}(f). Here, we resolve (far-field) interference for both condensate modes with extracted intensity profiles shown in Fig.~\ref{Fig4}(g) for ground state (red) and first excited state (blue). For comparison we plot the calculated un-modulated profiles (black-dashed) that resemble incoherent emission between both condensate traps. The experimentally observed far-field intensity modulation with destructive interference at $k_x=0$ reveals anti-phase synchronization between the two spatially separated condensates at both energy levels, i.e. $\Delta \theta_0 = \theta^{(2)}_0 -  \theta^{(1)}_0 = \pi $ and $\Delta \theta_1 = \theta^{(2)}_1 - \theta^{(1)}_1 = \pi$. One can easily show that phase-locking with equal phase-differences $\Delta \theta_0 = \Delta \theta_1$ of the coupled HO modes resembles in-phase synchronization of the two condensate oscillators as schematically illustrated for the two Bloch spheres in Fig.~\ref{Fig4}(a).

In the following, we describe the system's wave function in the tight-binding approach as a superposition of two spatially separated dual-mode condensates
\begin{equation}  \label{Eq.FourModeModel}
\Psi(\vect{r},t) = \sum_{j=1,2} \psi^{(j)}_0(t) \Psi^{(j)}_{00}(\vect{r}) + \psi^{(j)}_1(t) \Psi^{(j)}_{10}(\vect{r}).
\end{equation} 
 Synchronization of the two coupled condensate oscillators can be modeled by introducing a linear coupling $\mathcal{V}_{0,1}$ term between each of the trapped HO modes. In general this coupling term consists of both dissipative and non-dissipative parts, i.e. $\mathcal{V}_{0,1} = i \gamma_{0,1} + J_{0,1}$ ~\cite{PhysRevB.85.121301}. Using the short form notation $i\dot{\psi}^{(1,2)}_{0,1} = H^{(1,2)}_{0,1} \psi^{(1,2)}_{0,1}$ where $H_{0,1}$ is a function describing the two-level dynamics of an isolated condensate oscillator (i.e., square brackets of Eqs.~\eqref{Eq.complexEquations}) we can now write the 4 coupled equations
\begin{subequations}
\begin{align} 
i\dot{\psi}^{(1)}_{0} &= H^{(1)}_{0} \psi^{(1)}_{0} + \left( i \gamma_0 + J_0 \right) \psi^{(2)}_0,\\
i\dot{\psi}^{(2)}_{0} &= H^{(2)}_{0} \psi^{(2)}_{0} + \left( i \gamma_0 + J_0 \right) \psi^{(1)}_0,\\
i\dot{\psi}^{(1)}_{1} &= H^{(1)}_{1} \psi^{(1)}_{1} + \left( i \gamma_1 + J_1 \right) \psi^{(2)}_1,\\
i\dot{\psi}^{(2)}_{1} &= H^{(2)}_{1} \psi^{(2)}_{1} + \left( i \gamma_1 + J_1 \right) \psi^{(1)}_1,
\end{align} 
\end{subequations}
to describe the dynamics of each of the mode amplitudes $\psi^{(1,2)}_{0,1}$ constituting the total condensate wave function in Eq.~\eqref{Eq.FourModeModel}. It is known that dissipative coupling with $\gamma_{0,1} < 0$ ($\gamma_{0,1} > 0$) facilitates phase-locking of each trapped condensate mode with phase difference $\Delta \theta_{0,1} = \theta^{(2)}_{0,1} -  \theta^{(1)}_{0,1} = \pi (0)$~\cite{PhysRevB.85.121301}. Indeed, direct substitution of the in-phase or the anti-phase ansatz $\psi_{0,1}^{(1)} = \pm \psi_{0,1}^{(2)}$ respectively results in simply the same set of coupled equations as given by Eqs.~\eqref{Eq.complexEquations} with shifted $\omega_{0,1} \to \omega_{0,1} \pm J_{0,1}$ and $p_{0,1} \to p_{0,1} \pm \gamma_{0,1}$. A rigorous stability analysis of these and more exotic solutions is beyond the scope of the current study. While the far-field emission pattern of our experimentally realized coupled-trap system displayed in Fig.~\ref{Fig4}(f) indicates dissipative coupling with $\gamma_{0,1} < 0$, other synchronization patterns are expected when changing the separation distance between the two traps~\cite{PhysRevB.101.155402}.     

%%%%%%%%%%%%%%%%%%%%%%%%%%%%%%%%%%%%%%%%%
%%%%%%%%%%%%%%%%%%%%%%%%%%%%%%%%%%%%%%%%%
\section{Discussion}
In this work we have demonstrated optical generation, tunability and coupling of macroscopic two-level systems realized in trapped polariton condensates. Non-resonant excitation using annular beam profiles enables optical control over the mode-composition of each trapped condensate. As we have shown, dual-mode condensation comprising the two energetically-lowest trapped modes results in nearly terahertz (ps-timescale) spatial density oscillations in close analogy to the dynamics of a simple mechanical oscillator. Amplitude and frequency of such a condensate oscillator are both tuneable through modulation of the excitation laser beam. Dissipative coupling between two closely spaced condensate traps, each of which represents a (dual-mode) oscillator,  facilitates phase-locking of their spatial oscillations. This effect represents a quantum analogue of Huygen's clock synchronization realized by spatially oscillating coherent many-particle states. Our work demonstrates an interconnectable element for future polaritonic circuitry, and lays out directions for polaritonic applications utilizing coupled macroscopic two-level systems.

%%%%%%%%%%%%%%%%%%%%%%%%%%%%%%%%%%%%%%%%%
%%%%%%%%%%%%%%%%%%%%%%%%%%%%%%%%%%%%%%%%%
\section{Acknowledgments}
The authors acknowledge the support of the Skoltech NGP Program (Skoltech-MIT joint project), the UK’s Engineering and Physical Sciences Research Council (grant EP/M025330/1 on Hybrid Polaritonics), and the RFBR projects No. 20-52-12026 (jointly with DFG) and No. 20-02-00919. H.S. acknowledges hospitality provided by the University of Iceland.
%%%%%%%%%%%%%%%%%%%%%%%%%%%%%%%%%%%%%%%%%
%%%%%%%%%%%%%%%%%%%%%%%%%%%%%%%%%%%%%%%%%
\section{Methods} \label{section_Methods}
Experiments are conducted on a strain-compensated semiconductor microcavity with InGaAs quantum wells held in a cold finger cryostat at a temperature $T\approx 6\;\mathrm{K}$. The continuous wave Gaussian excitation laser (circular polarization, $\lambda \approx 785\;\mathrm{nm}$) is modulated using a spatial light modulator in the Fourier plane of the optical setup and focused with (elliptical) annular beam shape onto the microcavity using an $\mathrm{NA}=0.4$ microscope objective lens. The full-width-at-half-maximum of the laser profile in radial direction is $\approx 1.5\;\mathrm{\upmu m}$. An acousto-optic modulator transforms the continuous wave signal into square wave packets at $5\;\%$ duty cycle and $10\;\mathrm{kHz}$ repetition rate to avoid heating of the sample. We operate at a negative exciton-photon detuning of $\Delta \approx 5.5\;\mathrm{meV}$. Resulting polariton photoluminescence ($\lambda \approx 857\;\mathrm{nm}$) is collected in reflection geometry, filtered from the excitation laser using a long-pass filter and imaged simultaneously onto two different charge-coupled device (CCD) sensors. One sensor is placed at the output of a $750\;\mathrm{mm}$ spectrometer (equipped with a $1800\;\mathrm{grooves}/\mathrm{mm}$ grating) such that through combination of optical lenses we can choose to image spectrally-resolved momentum-space (far-field) or real-space (near-field) photoluminescence. The second sensor is placed at the output of a modified Mach-Zehnder interferometer, in which one mirror is replaced by a retro-reflector mounted on a translation stage. Here, we chose to image the interference of the real-space (near-field) polariton photoluminescence with its retro-reflected and time-shifted distribution, i.e. $|\Psi(\vect{r,t})+\exp{(i \vect{k} \vect{r})}\Psi(-\vect{r},t+\tau)|^2$. The wavevector $\vect{k}$ is controlled by transverse displacements of the retro-reflected beam and determines the orientation and periodicity of the interference fringes. Recorded interferograms are analyzed by means of off-axis digital holography to obtain magnitude and phase of the first-order correlation function $g^{(1)}(\vect{r},-\vect{r},\tau)$, where $\tau$ is the temporal shift controlled by the relative path-length difference of the two interferometer arms.   
%%%%%%%%%%%%%%%%%%%%%%%%%%%%%%%%%%%%%%%%%
%%%%%%%%%%%%%%%%%%%%%%%%%%%%%%%%%%%%%%%%%
\appendix
\renewcommand{\appendixname}{APPENDIX} %Capital letters
\renewcommand{\thesection}{\Alph{section}} %Subsections with roman capital letter
%%%%%%%%%%%%%%%%%%%%%%%%%%%%%%%%%%%%%%%%%
%%%%%%%%%%%%%%%%%%%%%%%%%%%%%%%%%%%%%%%%%
\section{COUPLED CONDENSATE MODES IN A TRAP} \label{Appendix_Theory}
The dynamics of the macroscopic wave function $\Psi(\vect{r},t)$ of the driven-dissipative polariton condensate can be described in mean field formalism~\cite{Wouters_PRL2007} as
\begin{equation} \label{Eq.GPE}
i \dot{\Psi} = \left[ \frac{-\hbar \nabla^2}{2m} + \alpha |\Psi|^2 -\frac{i\gamma_c}{2} + (g+i)n  \right] \Psi.
\end{equation} 
Here, intrinsic non-linearity and dissipation of polaritons are described with parameters $\alpha$ and $\gamma_c$. A pump induced background reservoir $n(\vect{r},t)$ provides both gain and a repulsive potential for the condensate wave function with interaction parameter $g>0$. In the following we assume that the reservoir can be removed from the dynamics of the condensate and only the simplest particle saturation mechanism is considered, i.e. $n(\vect{r},t) \approx P(\vect{r})(1-R |\Psi(\vect{r},t)|^2)$ where $R^{-1}$ is the saturation density. Assuming that the modulation to the reservoir from the condensate is small in the range of investigated pump powers and corresponding condensate densities, $1\gg R|\Psi(\vect{r},t)|^2$, the condensate trapping potential is approximately defined by the term $g P(\vect{r})$ where $P(\vect{r})$ denotes the constant, non-resonant and annular laser intensity profile. The eigenmodes of the two-dimensional trap satisfy
\begin{equation} \label{Eq.Eigenmodes}
\left[ \frac{-\hbar \nabla^2}{2m} + gP(\vect{r}) \right] \Psi_{nm}(\vect{r}) = \omega_{nm} \Psi_{nm}(\vect{r}).
\end{equation} 
and are taken to be normalized and real-valued. Since we operate in a regime, in which only the trap's ground state $\Psi_{00}$ and first excited state $\Psi_{10}$ are populated, we project the total wave function $\Psi$ onto a truncated Hilbert space 
\begin{equation} 
\Psi(\vect{r},t)=\psi_0(t)\Psi_{00}(\vect{r}) + \psi_1(t)\Psi_{10}(\vect{r}),
\end{equation} 
where $\braket{\Psi_{i0}|\Psi_{j0}}=\delta_{ij}$. Therefore, the problem can be described as an effective two-level system following the equations of motion
\begin{equation} 
\begin{aligned}[b]
i \dot{\psi}_{0,1} =  &\left[ \omega_{0,1} + i p_{0,1} + (\alpha_{0,1} - i \sigma_{0,1}) |\psi_{0,1}|^2 \right. \\
 & \left. +  (\beta - i \kappa) |\psi_{1,0}|^2  \right] \psi_{0,1}. 
\end{aligned}
\end{equation} 
Here, we have neglected nonlinear off-resonant mixing terms between the modes which vanish in the time average of our continuous-wave experiment. Each mode $j={1,2}$ is driven by its net gain $p_{j}=\bra{\Psi_{j0}} P \ket{\Psi_{j0}} - \gamma_c/2$, and stabilized by both its self-saturation term  $\sigma_j$ and cross-saturation term $\kappa$. These non-linear damping terms are determined by the overlap integrals
\begin{subequations}
\begin{align} 
\sigma_j &= \int P(\vect{r}) \Psi_{j0}^4(\vect{r})  \mathrm{d}\vect{r}, \\
\kappa &= 2 \int P(\vect{r}) \Psi_{00}^2(\vect{r})   \Psi_{10}^2(\vect{r}) \mathrm{d}\vect{r}.
\end{align} 
\end{subequations}
Nonlinear blue-shift of each mode due to Kerr and cross-Kerr effects are included in the integrals 
\begin{subequations}
\begin{align} 
\alpha_j &= \int \Psi_{j0}^4(\vect{r})  \mathrm{d}\vect{r}, \\
\beta &= 2 \int \Psi_{00}^2(\vect{r})   \Psi_{10}^2(\vect{r}) \mathrm{d}\vect{r}.
\end{align} 
\end{subequations}
\section{CONDENSATE LINEWIDTH} \label{Appendix_section_CoherenceTime}
\begin{figure}[!t]
	\center
	\includegraphics[]{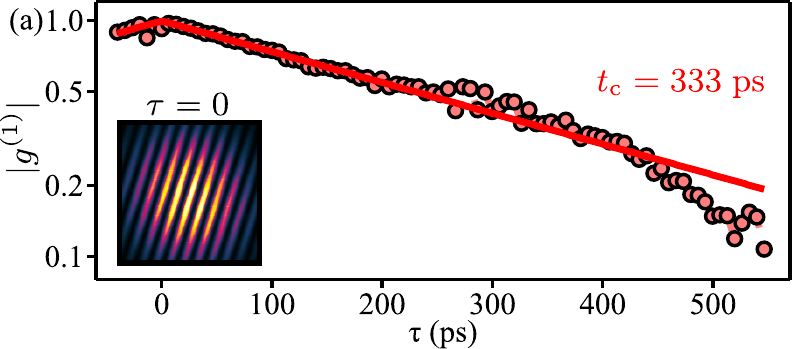}
	\caption{First-order coherence function $|g^{(1)}(\tau)|$ of the ground state $\Psi_{00}$ mode excited at pump power $P=2.1P_\mathrm{thr}$ in an elliptical trap. An exponential fit (red line) of the temporal decay of coherence yields a $1/e$ coherence time of $t_\mathrm{c}=333\;\mathrm{ps}$. Inset shows the measured interference of the condensates real-space emission with its retro-reflected version at $\tau=0$ time delay.}
	\label{Fig_Appendix_CoherenceTime}
\end{figure}
We investigate the linewidth of a single-mode trapped polariton condensate through analysis of the temporal auto-correlation function of its emission. We choose an elliptical annular pump with major axis $\approx 14.2\;\mathrm{\upmu m}$ and minor axis $\approx 10.6\;\mathrm{\upmu m}$ to excite a single-mode ground state $\Psi_{00}$ condensate above condensation threshold at $P=2.1P_\mathrm{thr}$. Real-space and energy-resolved photoluminescence of this condensate are presented in the main text in Figs.~\ref{Fig2}(c) and (f). The measured decay of temporal coherence $|g^{(1)}(\vect{r}=0, \vect{r}=0; \tau)|$ extracted at the Gaussian-shaped condensate center $\vect{r}=0$ is shown in Fig.~\ref{Fig_Appendix_CoherenceTime}. We find a single exponential decay of coherence with $1/e$ decay time of $t_\mathrm{c} = 333\;\mathrm{ps}$. Thus, for a Lorentzian lineshape we calculate a (full-width-at-half-maximum) condensate linewidth of $\Delta E = 2\hbar/\tau_{\mathrm{c}} \approx 4\;\mathrm{\upmu eV}$, which is below our experimental spectral resolution $\Delta \approx 0.02\;\mathrm{meV}$. 
%%%%%%%%%%%%%%%%%%%%%%%%%%%%%%%%%%%%%%%%%
%%%%%%%%%%%%%%%%%%%%%%%%%%%%%%%%%%%%%%%%%
\section{TUNING OF OSCILLATION FREQUENCY} \label{Appendix_section_FrequencyTuning}
\begin{figure}[b]
	\center
	\includegraphics[]{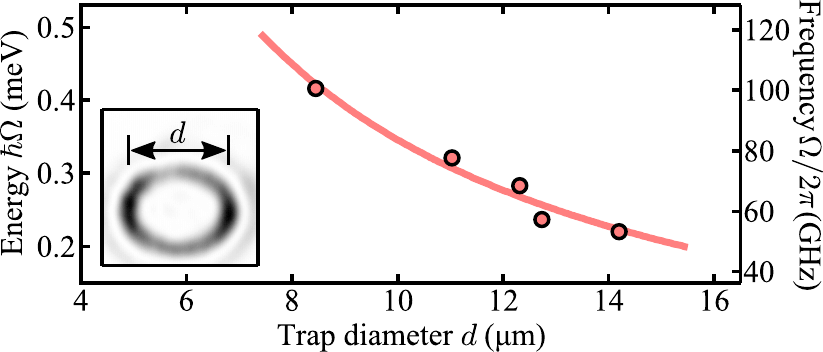}
	\caption{Measured energy splitting $\hbar \Omega$ between the lowest two modes $\Psi_{00}$ and $\Psi_{10}$ in an elliptical trap with diameter $d$ of the major axis. Inset depicts a typical pump laser profile. A power-law decay $\Omega \propto 1/d$ (red line) is shown as a guide for the eye.}
	\label{Fig_Appendix_TrapRadiusScan}
\end{figure}
Control over the oscillation frequency of the trapped condensate system is achieved by tuning of the energy splitting $\hbar\Omega = E_{10} - E_{00}$ between the two competing HO modes. A change in spatial dimension of the elliptical annular pump beam profile directly modifies the near-parabolic polaritonic potential landscape, and, thus, allows to control the HO mode energy splitting. In Fig.~\ref{Fig_Appendix_TrapRadiusScan} we show the measured energy splitting $\hbar \Omega$ (left) between $\Psi_{00}$ and $\Psi_{10}$ modes and the corresponding calculated oscillation frequency $(2\pi)^{-1}\Omega$ (right) for varying diameter $d$ of the elliptical trap pump profile along its (horizontal) major axis. In our experiment we simultaneously modify the vertical diameter of the ellipse to guarantee two-mode condensation of ground state $\Psi_{00}$ and excited state $\Psi_{10}$, i.e. we quench the population of the $\Psi_{01}$ mode. Hereby, the eccentricity $e$ of the elliptical pump profile stays the range of $\approx 0.6-0.8$ for all points shown in  Fig.~\ref{Fig_Appendix_TrapRadiusScan}. A typical excitation laser profile is illustrated in the inset of Fig.~\ref{Fig_Appendix_TrapRadiusScan}. We experimentally demonstrate tunability of the condensate's spatial oscillations in the frequency range of $50\;\mathrm{GHz} - 100\;\mathrm{GHz}$, which corresponds to oscillation periods of $20\;\mathrm{ps} - 10\;\mathrm{ps}$.

%%%%%%%%%%%%%%%%%%%%%%%%%%%%%%%%%%%%%%%%%
%%%%%%%%%%%%%%%%%%%%%%%%%%%%%%%%%%%%%%%%%
\bibliography{refs}
\end{document}